\documentclass[12pt,manuscript]{elsart}
\usepackage{epsfig,harvard,amssymb,rotating,lscape}

\makeatletter
\def\elsartstyle{%
        \def\normalsize{\@setfontsize\normalsize\@xiipt{14.5}}
        \def\small{\@setfontsize\small\@xipt{13.6}}
        \let\footnotesize=\small
        \def\large{\@setfontsize\large\@xivpt{18}}
        \def\Large{\@setfontsize\Large\@xviipt{22}}
        \skip\@mpfootins = 18\p@ \@plus 2\p@
        \normalsize
}
\makeatother

\def\ar{{Annu. Rev. Astron. Astrophys.} \,}
\def\aa{{Astron. Astrophys.} \,}
\def\apj{{Ap. J.} \,}
\def\apjs{{Ap. J. Supp.} \,}

\def\mn{{MNRAS} \,}
\def\na{{NewA} \,}

\def\phr{{Phys. Rep.} \,}
\def\ea{et al. \,}
\def\bibcode#1{(\texttt{#1})}
\def\eg{{e.g.\ }}

\def\pasj{{Publ. Astron. Soc. Japan} \,}
\def\casp{{Comments. Astrophys. Space Phys.}\,}
\begin{document}

\begin{frontmatter}
\title{Sunyaev-Zeldovich  Cluster Counts as a Probe of Intra-Supercluster Gas}

\author{Sharon Sadeh\thanksref{email}}
\address{School of Physics and Astronomy, Rayomnd and Beverly Sackler
Faculty of Exact Sciences, Tel Aviv University, Tel Aviv, 69978, Israel}

\thanks[email]{E-mail: shrs@post.tau.ac.il}

\and

\author{Yoel Rephaeli}
\address{School of Physics and Astronomy, Raymond and Beverly Sackler
Faculty of Exact Sciences, Tel Aviv University, Tel
Aviv, 69978, Israel, \\and\\ Center for Astrophysics and Space
Sciences, University of California, San Diego, La Jolla,
CA\,92093-0424}

\begin{abstract}

X-ray background surveys indicate the likely presence of diffuse warm 
gas in the Local Super Cluster (LSC), in accord with expectations from 
hydrodynamical simulations. We assess several other manifestations of 
warm LSC gas; these include anisotropy in the spatial pattern of cluster 
Sunyaev-Zeldovich (S-Z) counts, its impact on the CMB temperature power 
spectrum at the lowest multipoles, and implications on measurements of 
the S-Z effect in and around the Virgo cluster.

\end{abstract}

\end{frontmatter}

\section{Introduction}

As much as $30-40\%$ of all baryons are believed to be in warm gas in 
large scale filamentary structures connecting clusters and superclusters 
(SCs) of galaxies. Observational evidence for the so-called \emph{warm-hot 
intergalactic medium} (WHIM) is still quite rudimentary. First systematic 
attempts to detect the low X-ray surface brightness enhancement that could 
possibly be expected from the WHIM were conducted by Persic \ea (1988, 1990) 
who searched the HEAO-1 A2 database in directions to a sample of SCs. In 
both these analyses, as well as a similar analysis of Ginga observations 
of the Coma-A1367 SC (Tawara \ea 1993), only upper limits were obtained on 
the flux from the putative intra-SC (ISC) gas. More recently, evidence for 
soft X-ray excess emission in regions around several clusters has been 
claimed by Kaastra et al. (2003), who carried out a detailed analysis of 
XMM-Newton line and continuum measurements of extended regions around 
several clusters.

X-ray emission may be expected, of course, also from the WHIM in the Local 
SC (LSC); analysis of the HEAO-1 A2 database led Boughn (1999) to conclude 
that this emission was actually detected. Adopting a simple model for the 
morphology of the LSC, and assuming uniform temperature and density LSC 
gas, Boughn deduced the mean electron density to be                   
$n_e=2.5\cdot 10^{-6}\,\left(a/20\,Mpc\right)^{-1/2}\left
(kT_e/10\,keV\right)^{-1/4}\,cm^{-3}$, when using the specified scalings 
for the values of the electron temperature, $kT_e$, and the length of 
the semi-major axis of the LSC, $a$. More generally, Kaastra et 
al. (2003) have reported the detection of line and continuum WHIM 
emission in XMM-Newton observations of extended regions around five 
cluster. 
While the observational results on the properties of LSC gas are quite 
uncertain, gaseous filamentary structures seem to be a ubiquitous feature 
of the large scale mass distribution described by hydrodynamical 
simulations (e.g. Jenkins \ea 1998). Of considerable interest are the 
statistics and morphologies of these filaments; these are currently 
being quantified (\eg Colberg \ea 2004).

The scant observational information on the WHIM and its properties in the 
LSC provides strong motivation for assessing the feasibility of probing 
it by measurements of the Sunyaev-Zeldovich (S-Z) effect (Sunyaev         
\& Zeldovich 1972). The effect - a spectral change of the Planck spectrum 
due to Compton scattering of the cosmic microwave background (CMB) by 
electrons in clusters of galaxies - is a major cluster and cosmological 
probe (as reviewed by Rephaeli 1995, Birkinshaw 1999, and by Carlstrom \ea 
2002). In fact, the possibility of detectable S-Z signals in directions 
to SCs has been considered long ago (Persic, Rephaeli \& Boldt 1988, 
Rephaeli \& Persic 1992, Rephaeli 1993), but given the substantial 
uncertainty in the integrated pressure of ISC, no definite predictions 
could be made. Measurement capabilities have greatly advanced since then, 
so the prospects of detecting weak S-Z signals have correspondingly 
improved. 

Interest in the impact of ISC gas on the CMB has also increased. In 
particular, it has been suggested that gas in the Local Group may be 
responsible for the power suppression of primary CMB temperature 
anisotropy on large angular scales. This suppression was deduced from 
analyses of the COBE/DMR (Hinshaw \ea 1996) and WMAP databases (Bennett 
\ea 2003). However, the estimated low Comptonization parameter of LG 
gas makes this suggestion quite unlikely (Rasmussen \& Pedersen 2001). 
More recently, Abramo \& Sodr\'{e} (2003) proposed that warm LSC gas 
may be responsible for the deduced CMB low multipole power suppression, 
presumably caused by coincidental alignment of a hot spot in the CMB 
with the line of sight connecting the Galaxy with the Virgo cluster. 
At the relevant spectral range of the DMR and WMAP experiments ($\sim 
20-90\, GHz$), the S-Z diminution lowers the power in the low multipoles. 
Clearly, for this to occur the integrated electron pressure has to be 
sufficiently high.                                      

WHIM in the LSC may significantly screen our view of the CMB and S-Z
sky. In addition to the possible impact on the power spectrum of CMB
primary anisotropy, it will also affect the ability to measure the S-Z
induced anisotropy and cluster number counts. WHIM in the LSC may
appreciably increase S-Z cluster number counts. Moreover, the marked
ellipsoidal shape of the LSC, and the far off-center position of the
Galaxy, may result in a substantially anisotropic distribution of S-Z
counts across the sky. This anisotropy in S-Z counts is obviously in
addition to that generated by our motion in the CMB frame (the CMB
kinematic dipole) whose effect on cluster counts was recently explored 
by Chluba \ea (2004).

This paper is arranged as follows: the LSC gas model, the method for 
calculating the directional distribution of S-Z cluster counts, and 
the relevant expressions for calculating the S-Z angular power spectrum, 
are described in section~\S\,2. The main results of the calculations 
are presented in \S\,3. Additional aspects of the LSC gas model 
pertaining to S-Z measurements towards the Virgo cluster are 
discussed in \S\,4. A general discussion follows in \S\,5.

\section{Model and Method of Calculation}

The physical properties of ISC have not yet been well determined; what 
can be done at present is an attempt to incorporate the established 
observational results in a reasonable model in order to provide useful 
insight to guide upcoming observations, mostly of the S-Z effect. 
Relevant LSC properties that seem to have been {\it roughly} determined 
are the total mass, baryonic mass fraction, and its overall configuration. 
Knowing the baryon fraction in clusters (\eg Carlstrom \ea 2002), and 
adopting the scaling to obtain the global baryonic fraction (based on 
hydrodynamical simulations; Evrard 1997), $\sim 10-14\%$, the total LSC 
mass, $\sim 10^{15}\,M_{\odot}$, then yields an estimate for the LSC WHIM 
mass, $\sim 3-5.6\cdot 10^{13}\,M_{\odot}$, assuming (based on current 
expectations) that it comprises some $30-40\%$ of the total mass. The 
distribution of galaxies and clusters in the LSC can be described by a 
spheroid of semi-axes measuring $A=20\, Mpc$, $B=6.7\,Mpc$, $C=3.3\,Mpc$ 
(Tully 1982); the Galaxy is on the major axis of the LSC at a distance 
of $15\,Mpc$ from the center of the spheroid, where the Virgo cluster 
is located. 

Gas in the LSC is likely to be differently distributed than the galaxies, 
judging by the irregular filamentary structures seen in hydrodynamical 
simulations. However, it is reasonable to expect that most of the WHIM 
is concentrated along the major axis of the LSC mass distribution. Our 
specific model for gas in the LSC is based on the findings of Colberg 
\ea (2004), who analyzed results of N-body simulations (by Kauffmann 
\ea 1999) in order to characterize morphologically the filaments observed 
in the simulations. They found that most filaments have cross sections
of $1-2\,Mpc$, with the gas density decreasing outward from the 
major axis of the filament as $\sim r^{-2}$ beyond some scale radius 
$r_s$. The LSC gas is assumed to have an ellipsoidal configuration with 
semi-axes (in Mpc) of either $20\times 1\times 1$ or $20\times 2\times 2$. 
The highly prolate ellipsoid represents the gas `filamentary' structure. 
The gas mean density in these structures are $n_e = 2.4\cdot 10^{-5}\,
cm^{-3}$, or a factor $4$ lower, respectively; its temperature is 
scaled to a value of $1\,keV$).

Clearly, the expected anisotropic morphology of ISC gas in the LSC will be 
reflected in anisotropic cluster counts when S-Z surveys are conducted in 
different sky directions. The Comptonization parameter towards the Virgo 
cluster, a direction which in spherical coordinates corresponds to angles 
$(\phi,\theta)=(0,\pi/2)$, is 
\begin{equation}
y=\int\sigma_T\frac{k T_e}{m_e c^{2}}\,n_e\, d\ell 
= 3.4\cdot 10^{-6}\left(\frac{kT_e}{1 \, keV}\right),
\end{equation}
whereas the corresponding value calculated for the opposite direction
$(\pi,\pi/2)$ is seven times lower. 
The non-negligible level of the Comptonization parameter of LSC gas may 
modify the observed distribution of S-Z clusters across the sky; the 
effective y-parameter measured along a los to a cluster will have 
contributions from both intracluster (IC) and ISC gas, giving rise to an 
intensification of the apparent S-Z signal due to the cluster alone. In 
spherical coordinates, the path length through the ISC gas configuration 
as a function of position angle $\hat{n}\equiv(\phi,\theta)$ is 
\begin{equation}
r(\phi,\theta)=\frac{\frac{30\sin{\theta}\cos{\phi}}{A^2}+
\sqrt{\left(\frac{30\sin{\theta}\cos{\phi}}{A^{2}}\right)^{2}-4\left(\frac{15^2}{A^2}-1\right)
\left[\frac{\sin^{2}{\theta}\cos^{2}{\phi}}{A^{2}}+\frac{\sin^{2}{\theta}\sin^{2}{\phi}}
{B^{2}}+\frac{\cos^{2}{\theta}}{C^{2}}\right]}}
{2\left[\frac{\sin^{2}{\theta}\cos^{2}{\phi}}{A^{2}}+
\frac{\sin^{2}{\theta}\cos^{2}{\phi}}{B^{2}}+\frac{\cos^{2}{\theta}}{C^{2}}\right]}.
\label{eq:rtf}
\end{equation}
The Comptonization parameter measured along this direction is then simply
\begin{equation}
y(\phi,\theta)=\sigma_T\frac{k T_{SC}}{m_e c^{2}}\cdot n_{SC}\cdot 
r(\phi,\theta).
\label{eq:ypar}
\end{equation}
where $T_{SC}$ and $n_{SC}$ denote the uniform temperature and electron 
density of the LSC gas.

The flux received from a cluster lying in the direction $(\phi,\theta)$ 
is a sum of the cluster flux and the S-Z signal generated in the LSC; 
assuming a $\beta$-King profile for the IC gas density with $\beta=2/3$ 
and an isothermal temperature distribution, the total flux at a frequency 
$\nu$ is
\begin{eqnarray}
& &\Delta F(\phi,\theta)\equiv\Delta F_{C}(\phi,\theta)+\Delta F_{SC}(\phi,
\theta)= \nonumber \\
\nonumber \\
& &\frac{2\left(k T_{\gamma}\right)^{3}}{\left(hc\right)^{2}}
\cdot\left(\frac{k \sigma_T}{m_e c^{2}}\right)\cdot g(x)
\left[2 n_0 T_0
r_c\int_0^{p\theta_c}\frac{\tan^{-1}{\sqrt{\frac{p^2-\left(\theta/\theta_c\right)^2}
{1+\left(\theta/\theta_c\right)^{2}}}}}{\sqrt{1+\left(\theta/\theta_c\right)^2}}\,d\,\Omega\right.
\\
\nonumber \\
& &+\left. n_{SC}T_{SC}\int_{0}^{2\pi}\int_{0}^
{\sigma_b}\,r(\phi,\theta)\,\theta\,d\theta\,d\phi\right], \nonumber 
\label{eq:totalf}
\end{eqnarray}
where $T_{cmb}=2.726^{\circ} K$, $T_0$, $n_0$, $T_{SC}$ and $n_{SC}$ 
denote the temperature and density of the cluster and LSC gas, 
respectively, $p\equiv R_{vir}/r_c$, $\sigma_b$ is the beam width, and 
$g(x)$ is the non-relativistic spectral form of the thermal S-Z effect, 
valid here owing to the relatively low temperatures of both the ISC and 
Virgo IC gas. At higher temperatures a relativistically correct 
expression (e.g. Shimon \& Rephaeli 2004) must be taken into account.

The calculations were carried out in a standard $\Lambda$CDM model, 
specified by the parameters $\Omega_m=0.3, \Omega_{\Lambda}=0.7, h=0.7, 
n=1, \sigma_8=1$. We use the cluster mass function of Sheth \& Tormen 
(2001) mass, and the temperature-mass relation 
\begin{equation}
T=1.3\left(\frac{M}{10^{15}\,h^{-1}\,M_{\odot}}\right)^{2/3}
\,\left(\Delta_c\,E^{2}\right)^{1/3}\,\left(1-2\frac{\Omega_{\Lambda}(z)}{\Delta_c}\right)\,keV,
\end{equation}
where $E^{2}=\Omega_m(1+z)^{3}+\Omega_{\Lambda}$. A non-evolving gas 
fraction $f_0=0.1$ is assumed. And since our treatment here is most 
relevant for the Planck all sky survey, we have selected the spectral 
band to be the HFI $\nu_0$=353 GHz, with $\Delta\nu=116\, GHz$; 
correspondingly, the beam profile is taken to be Gaussian with FWHM 
of $7.1'$.

\begin{figure}[ht]
\centering
\epsfig{file=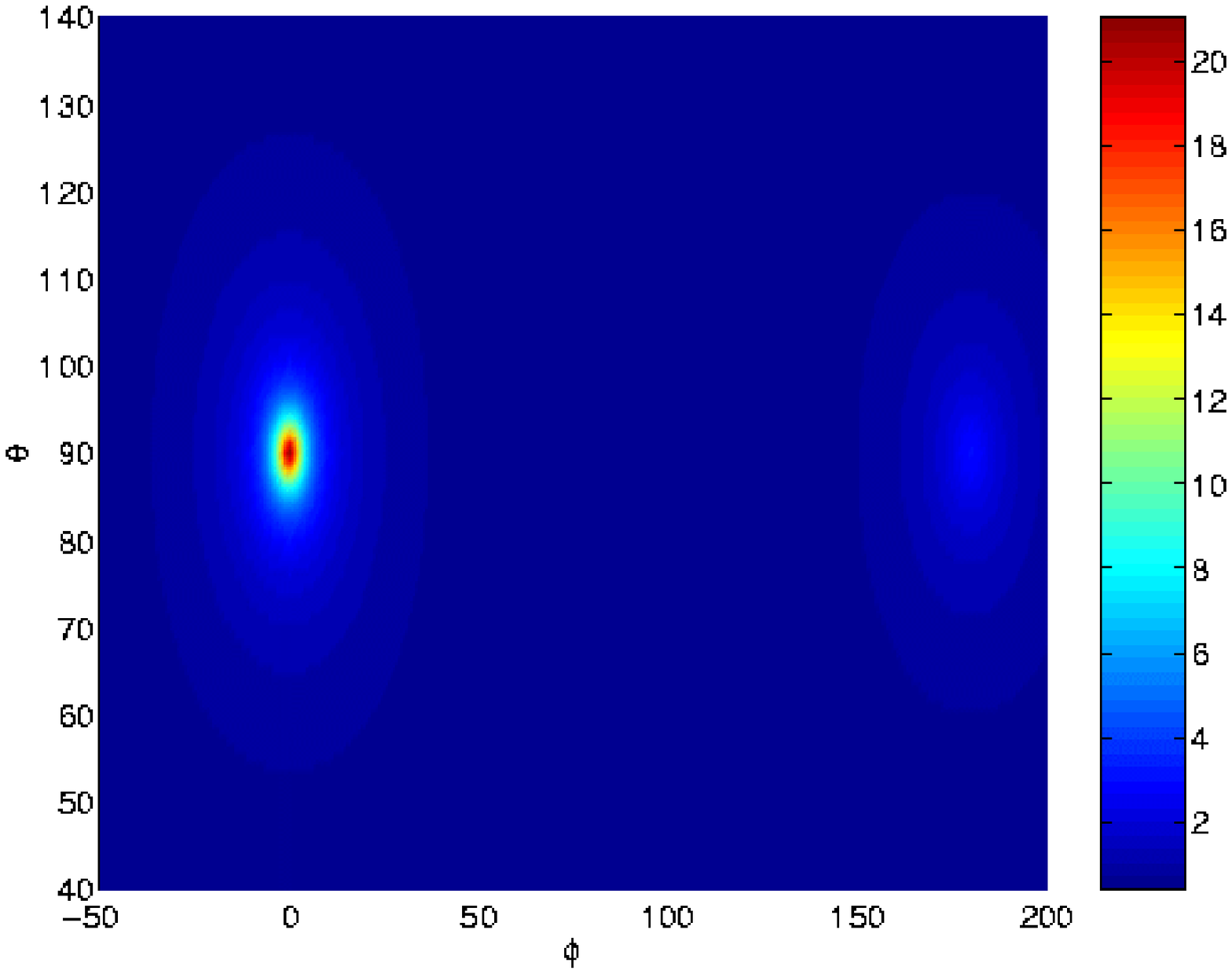, height=8.5cm, width=6.7cm,clip=}
\epsfig{file=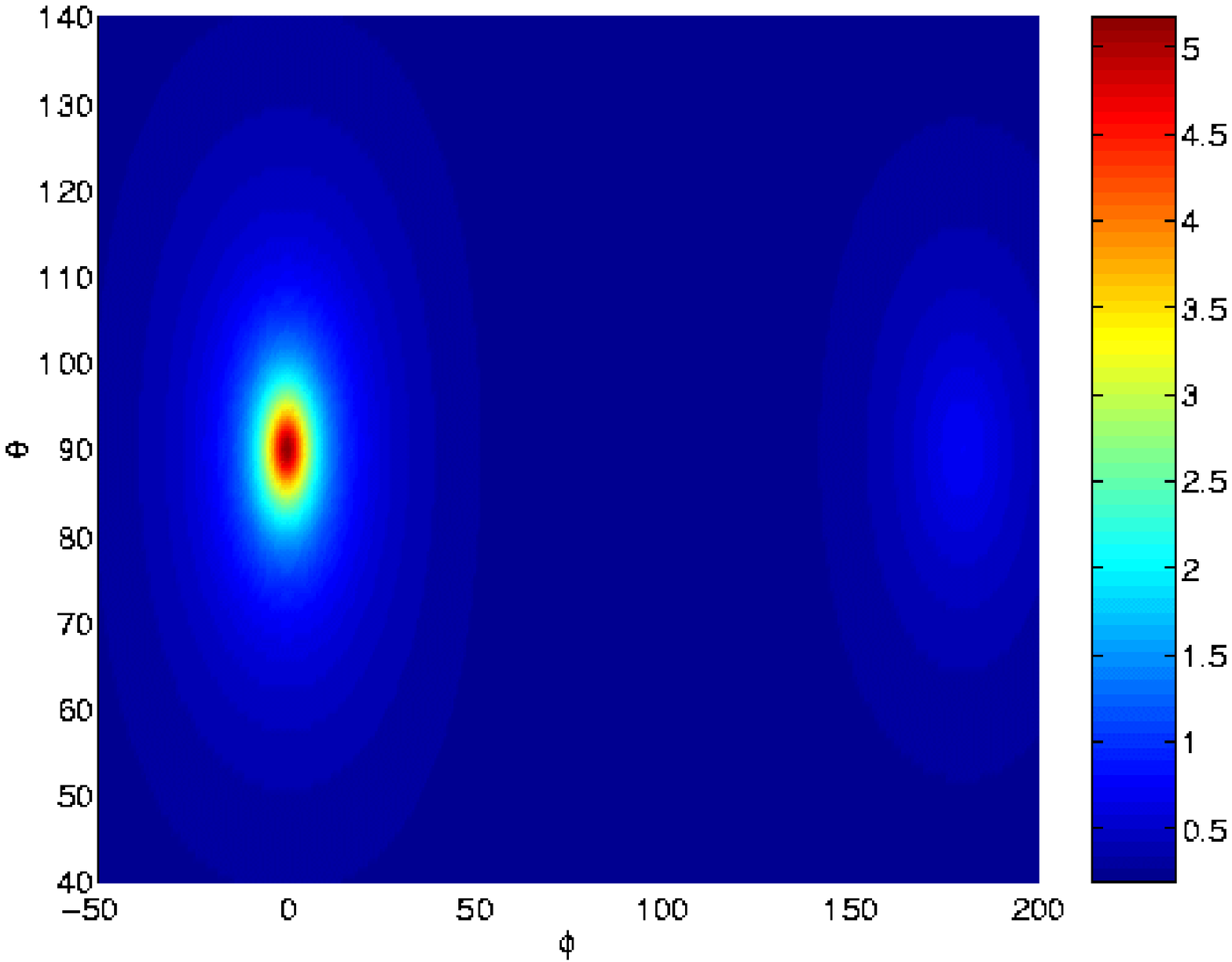, height=8.5cm, width=6.7cm,clip=}
\epsfig{file=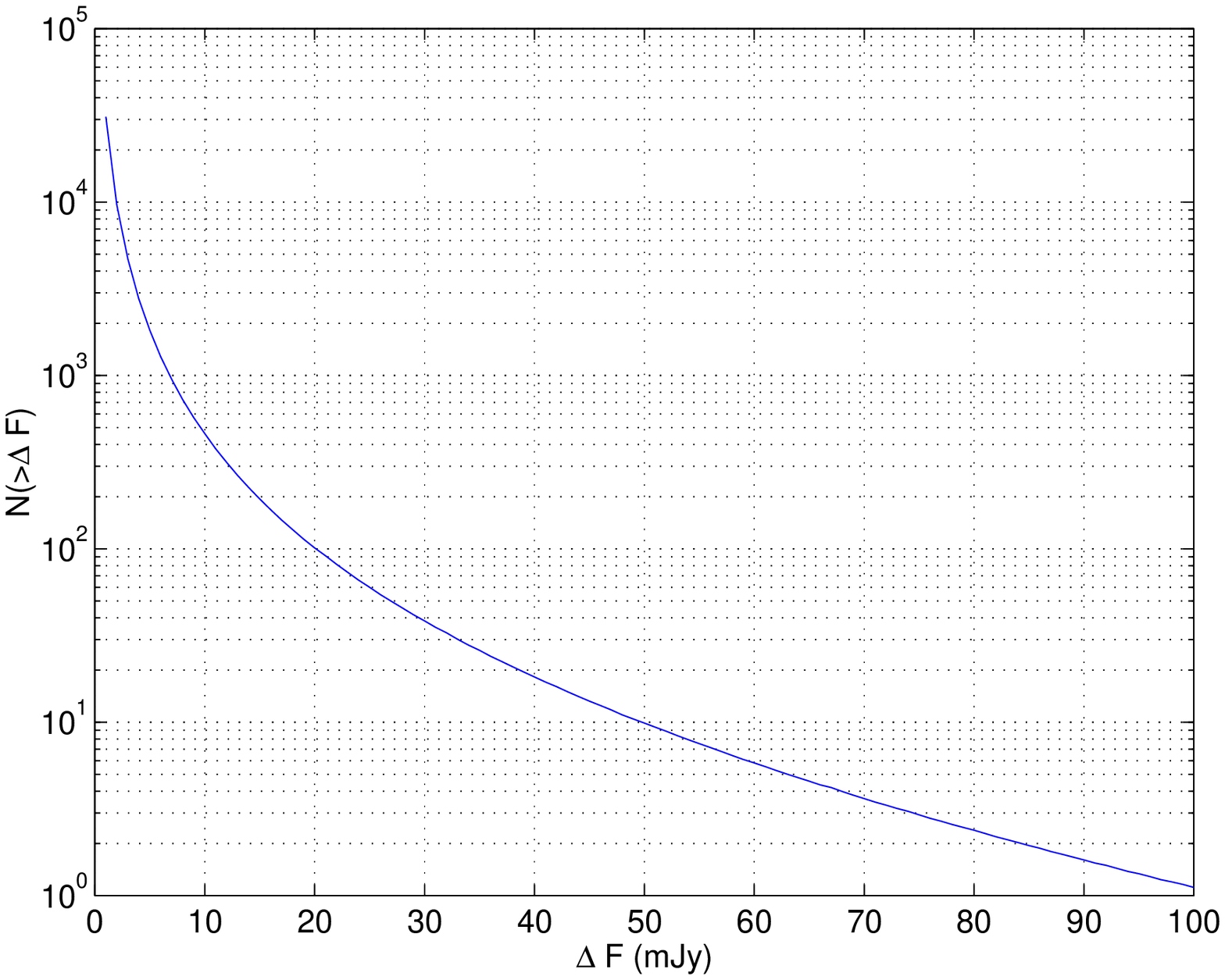, height=8.5cm, width=6.7cm,clip=}
\caption{Upper left- and right-hand panels: color-coded flux levels 
of the S-Z signal at 353 GHz (in units of mJy) from LSC gas at a 
density $n_e = 2.4\cdot 10^{-5}\,cm^{-3}$, and a density which is a 
factor 4 lower, respectively. Lower panel: number of clusters within
  $5^{\circ}\times 5^{\circ}$ patches of sky, above flux limit $\Delta
  F_{\nu}$.}
\label{fig:lscmap}
\end{figure}

Number counts are calculated based on the expression
\begin{equation}
N(>\Delta\overline{F}_{\nu})=\int\,r^{2}\frac{dr}{dz}\,dz
\int_{\Delta\overline{F}_{\nu}(M,z)}\,N(M,z)\,dM,
\label{eq:numcounts}
\end{equation}
where the lower limit is assigned such that the flux received from a 
cluster with mass $M$ situated at redshift $z$ exceeds the flux limit 
of the survey. In order to estimate the number of clusters that may be 
observed above a flux limit within the framework of the LSC model, the 
LSC S-Z flux is mapped using the second term in equation~(4); the 
results are illustrated in the upper left-(model 1) and right-hand 
(model 2) panels of figure~\ref{fig:lscmap}. The number of clusters 
with flux exceeding the limit $\Delta F_{\nu}$ are then calculated 
using equation~(\ref{eq:numcounts}). Calculated number counts are then 
multiplied by $\sim 6\cdot 10^{-4}$ to yield the number of clusters 
expected within square patches of sky measuring 
$5^{\circ}\times 5^{\circ}$. These are depicted in the lower panel of 
figure~\ref{fig:lscmap}. For the detection of a cluster with a given 
flux to be possible, the combined S-Z flux from the cluster and the 
intervening LSC gas must exceed that of the measurement sensitivity. 
For example, if the latter is 30 mJy, then in a sky region where the 
S-Z flux due to the LSC is 20 mJy, the minimum detection level for a 
cluster is 10 mJy. Having determined the minimum flux, the expected 
cluster counts within a 25 square degree patch of sky in the requested 
direction can be read off the graph in the lower panel of 
figure~(\ref{fig:lscmap}).

The angular power spectrum of the S-Z effect due to LSC gas can be 
calculated using the usual spherical harmonic expansion of the 
temperature anisotropy. Here, however, the flat sky approximation 
is invalid due to the large angular scales involved. The basic 
expression is 
\begin{equation}
a_{\ell m}=\int_{0}^{2\pi}\int_{0}^{\pi}\frac{\Delta T}{T}(\phi,\theta)\,
y^{*}_{\ell m}(\phi,\theta)\sin{\theta}\,d\theta\,d\phi,
\end{equation}
which, after projecting the three dimensional function $\Delta T/T
(r,\phi,\theta)$ onto the two dimensional celestial sphere, and some
algebraic manipulation, assumes the form
\begin{equation}
a_{\ell m}=\frac{2\sigma_T k_B T_e n_e}{m_e
c^{2}}\int_{0}^{2\pi}\int_{0}^{\pi}r(\phi,\theta)\cdot y^{*}_{\ell
m}(\phi,\theta)\sin{\theta}\,d\theta\,d\phi,
\end{equation}
where $r(\phi,\theta)$ is given by equation~(\ref{eq:rtf}). 
The angular power spectrum may then be easily calculated as
\begin{equation}
\ell(\ell+1)/2\pi\cdot C_{\ell}=\ell(\ell+1)/2\pi\cdot\frac{\sum_{m=-\ell}^{\ell}
\left|a_{\ell m}\right|^{2}}{2\ell+1}.
\end{equation}

\section{Results}

For flux limits of $30\,mJy$ and $60\,mJy$ the corresponding all-sky 
cluster counts sum up to $\sim 40\,000$ and $\sim 4\,300$, respectively. 
Assuming that the cluster population is homogeneously distributed across 
the sky, this amounts to $\sim 24$ and $\sim 3$ clusters in each 25 
square degrees (rectangular sky region). It is important to emphasize 
that the number of potentially detected clusters in future S-Z surveys 
predicted by numerical studies is quite sensitive to the choice of 
parameters characterizing the background cosmology and cluster 
properties, and consequently, these counts merely reflect the specific 
modeling detailed in the last section. Now suppose that observations are 
conducted in the directions
$\sim(\phi,\theta)=(0^{\circ},90^{\circ})$ (i.e. towards the longest
possible los within the LSC halo), $(5^{\circ},90^{\circ})$,
$(180^{\circ},90^{\circ})$ (designating the direction opposite to the 
Virgo cluster), and, e.g., $(180^{\circ},20^{\circ})$. The S-Z fluxes 
induced by the LSC gas in these directions are (see 
figure~\ref{fig:lscmap}) $\sim 21, 7, 3, 0.5\,mJy$, respectively, 
implying that for a flux detection limit of $30\,mJy$ clusters will 
be detected if their S-Z signals exceed the complementary values 
($9, 23, 27, 29.5\,mJy$, respectively) in these directions. The 
corresponding limiting values for a mean gas density of $n_e = 
6\cdot 10^{-6}\,cm^{-3}$ are $\sim 25,27,29.3,29.7\, mJy$, respectively. 

The number of clusters that could be detected in these directions, i.e., 
those which meet the detectability criterion, can be easily read off the 
diagram in figure~\ref{fig:lscmap}. For example, an experiment with a 
detection limit of $30\,mJy$ yields $\sim 570, 73, 50, 40$ clusters per 
25 square degrees. The corresponding counts for the larger ellipsoid 
(with gas density lower by a factor of 4) are $\sim 60, 50, 40, 40$. 
A limiting flux of $60\,mJy$ yields counts of $\sim 20, 8, 7, 6$ in 
the first case, and $\sim 6-8$ for all directions in the second case. 
While these estimates of the cluster number counts are substantially 
uncertain, and the fact that there is also considerable variation 
along different los across 25 square degrees (particularly towards 
both ends of the spheroid major axis), the results nonetheless 
demonstrate the possible impact of LSC gas, especially in a survey with 
$30\,mJy$ flux limit, and in direction to the Virgo cluster. The 
difference between counts along los in the Virgo region and those in 
other directions markedly exceed the statistical uncertainty (of a 
purely Poissonian distribution of clusters). 

\begin{figure}[ht]
\centering
\epsfig{file=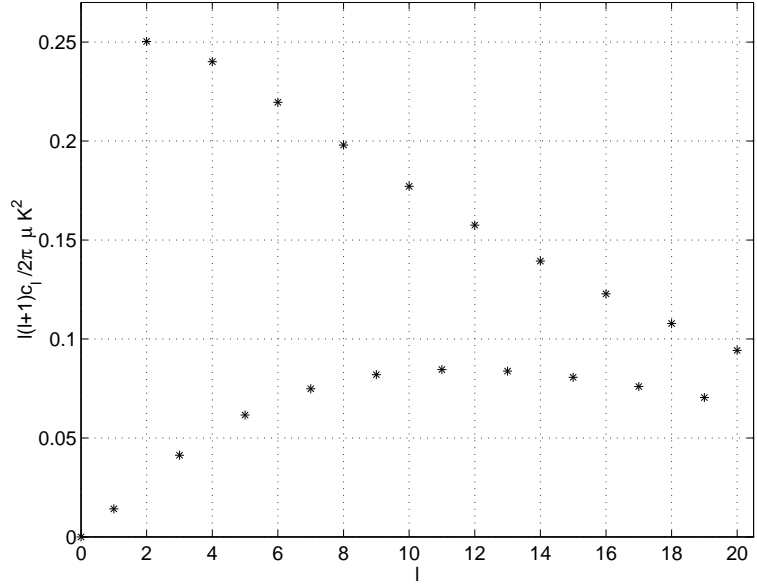, height=8cm, width=10cm,clip=}
\caption{The angular power spectrum of the S-Z effect induced by the
 gaseous LSC halo.}
\label{fig:szps}
\end{figure}

The angular power spectrum of the S-Z effect induced by LSC gas 
is illustrated in figure~\ref{fig:szps}. The maximum power level is 
significantly lower than the corresponding values found by Abramo \&
Sodr\'{e}. This is due to our lower values of the density and 
temperature, each lower by a factor of $\sim 2$ than their values 
($5\cdot 10^{-5}\,cm^{-3}$, and 2 keV), and the very different 
LSC gas configuration assumed by Abramo \& Sodr\'{e}, which is
reflected in a different distribution of power among the explored
multipole range. In particular, more power is seen at higher
multipoles (with regard to Abramo \& Sodr\'{e}'s results), which is
expected owing to its highly prolate morphology. Consequently, 
given the substantial uncertainty in the properties of ISC gas, 
one cannot convincingly argue that the observed suppression of the 
primary CMB temperature power spectrum is due to Compton scattering in 
the LSC. 

\section{Implications for the Virgo Cluster}

The central location of Virgo in the LSC implies that measurement 
of the S-Z effect in this cluster will likely include a substantial 
contribution from LSC gas. The Comptonization parameter measured along 
the direction $(\phi,\theta)$ is 
\begin{equation}
y=\sigma_T\frac{kT_{SC}}{m_ec^{2}}n_{SC}\,r(\phi,\theta)+
2\sigma_T\frac{kT_0}{m_ec^{2}}n_0\,r_c\,\ell(\phi,\theta),
\end{equation}
where $r(\phi,\theta)$ is given in equation~(\ref{eq:rtf}) and 
\begin{equation}
\ell(\phi,\theta)=\left[1+\left(\frac{\theta}{\theta_c}\right)^{2}\right]^{-\delta}\cdot
\sqrt{p^2-(\theta/\theta_c)^{2}}\cdot
H_2F_1\left[\frac{1}{2},\delta,\frac{3}{2},\frac{-(p^2-(\theta/\theta_c)^2)}
{1+(\theta/\theta_c)^2}\right],
\end{equation}
is the integrated line of sight through the Virgo cluster. 
Here $\theta_c$ is the core radius, $\delta\equiv 3/2\beta\gamma$
where $\gamma$ is a polytropic index, $p$ is the virial to core radius
ratio, and $H_2F_1$ is the hypergeometric function. The main 
gas configuration in Virgo is centered on the giant elliptical M87, 
to which we refer here. Excluding a relatively small region in the 
center of M87, the gas is taken to be isothermal with a (King) $\beta$ 
density profile, having (Matsumoto \ea 2000)
$n_0=0.019\,cm^{-3}$, $T_0=2.28\,keV$, $r_c=0.014\,Mpc$, $p=125$, and
$\beta=0.4$. 
The magnitude of $y$ along the central Virgo los increases by $\sim 20\%$ 
due LSC gas, its gradient across the cluster is appreciably shallower than 
would be expected from the intrinsic density profile, and the cluster 
seems slightly larger due to the enhancement of the cluster 
Comptonization parameter.

\begin{figure}[ht]
\centering
\epsfig{file=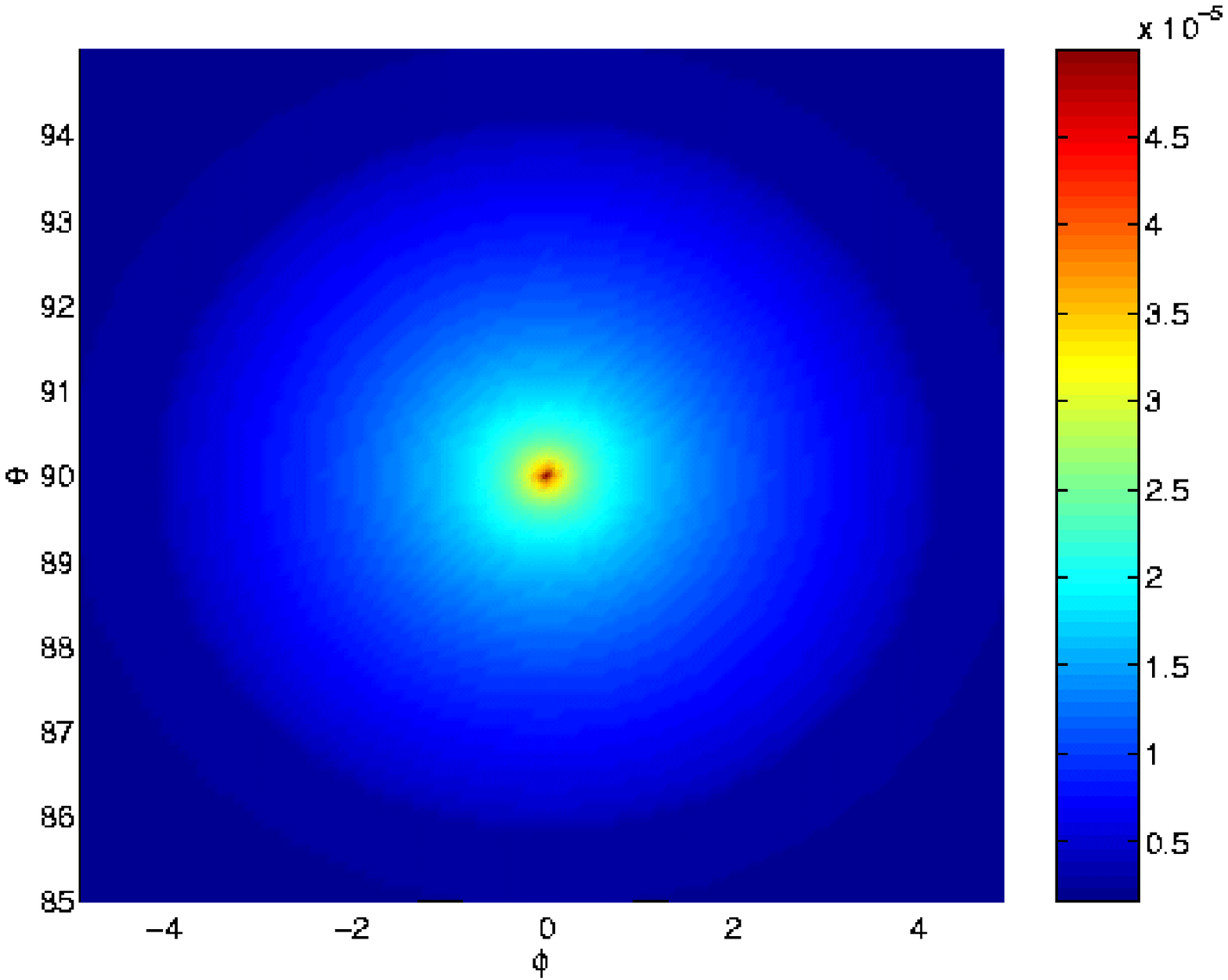, height=8.5cm, width=6.5cm, clip=}
\epsfig{file=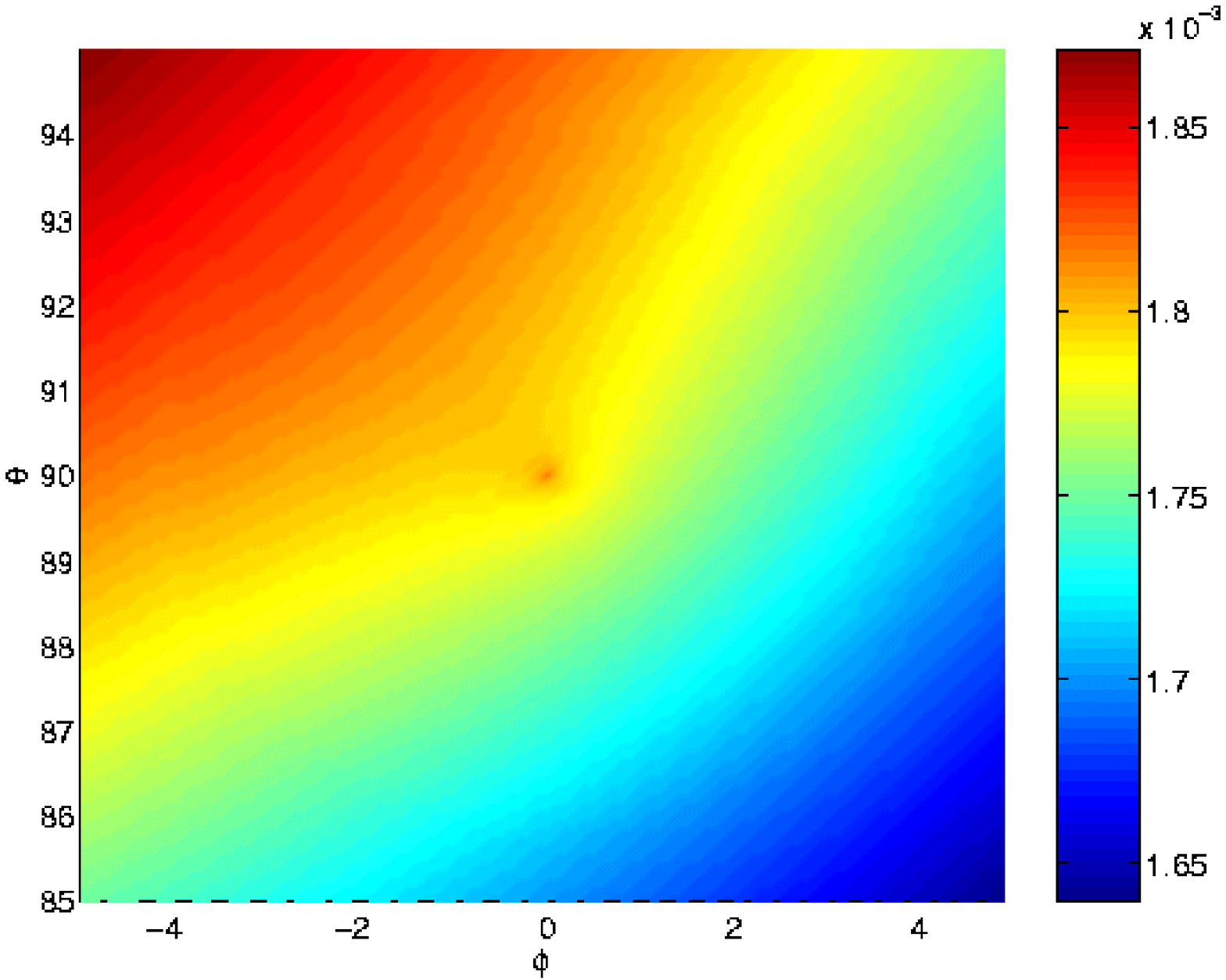, height=8.5cm, width=6.5cm, clip=}
\caption{The distribution of the relative temperature change around
  the Virgo cluster due to IC and ISC gas (left-hand panel), and combined
  with the temperature change induced by the CMB dipole term
  (right-hand panel).}
\label{fig:dip}
\end{figure}

The combined motion of the earth towards Virgo and the motion of the
LSC in the CMB frame accounts for the dipole term; the deduced velocity 
is $\sim 600\,km s^{-1}$ in the direction $(\ell=264^{\circ},
b=48^{\circ})$ in galactic coordinates (Fixen et al. 1996). The 
temperature change generated by the dipole in the direction of Virgo is
$\Delta T/T\sim 1.8\cdot 10^{-3}$, whereas the combined S-Z temperature 
change due to the LSC halo and Virgo cluster peaks at $\Delta T/T\sim 
5\cdot 10^{-5}$. Clearly, the dipole may interfere with S-Z measurements 
of the Virgo cluster and should therefore be taken into account. This 
would generally be the case for large angular diameter clusters, across 
which the dipole gradient can be significant. The temperature change due 
to the dipole term is simply 
\begin{equation}
\left(\frac{\Delta T}{T}\right)_{dipole}\equiv\frac{T_{obs}-T_0}{T_0}=
\frac{v}{c}\cos{\theta},
\end{equation}
whereas the temperature change due to the thermal S-Z effect in the R-J 
region is $\Delta T /T = s(x)\cdot y$, where $s(x)=x/\tanh(x/2)-4$. In 
order to be able to compare the contributions of both effects to the 
temperature change, the galactic coordinates specified above are 
converted into equatorial coordinates, and transformed such that the 
Virgo cluster lies in the direction $(\phi=0^{\circ},\theta=90^{\circ})$. 
In this coordinate system the velocity vector points at 
$\sim(\phi=340^{\circ},\theta=110^{\circ})$. Figure~\ref{fig:dip} 
illustrates the relative temperature change $(\Delta T/T)$ due to the 
LSC+Virgo S-Z signal, and the latter combined with the CMB dipole 
$(\Delta T/T)_{dipole}+(\Delta T/T)_{SC+Virgo}$ within a 
$10^{\circ}\times 10^{\circ}$ patch of sky centered around the Virgo 
cluster. The relative temperature change due to the dipole does not 
change considerably across the patch due to its small angular size and 
the fact that the angle formed between the dipole and the vector 
pointing at Virgo lies within the range $21^{\circ}\le\gamma\le 
35^{\circ}$, at a significant angular distance from the direction at 
which the cosine term of the dipole has the steepest slope, 
$\gamma=90^{\circ}$. The angle between the two vectors can be 
calculated using
\begin{equation}
\cos{\gamma}=\cos{\theta_1}\cos{\theta_2}+\sin{\theta_1}\sin{\theta_2}\cos{(\phi_1-\phi_2)}.
\end{equation}
The combined relative temperature change is still dominated by the 
dipole term, although its profile is now deformed due to the S-Z 
induced temperature change. In particular a 'warm spot' may be seen in 
the direction of Virgo (which translates into an enhancement of 
$\sim 3\%$ of the temperature change due to the dipole term in this 
direction), by virtue of the fact that at a frequency of 353 GHz the 
cluster behaves as a source for CMB photons. Consequently, it is obvious 
that the dipole term must be taken into account in measurements of S-Z 
signals in nearby clusters. 

The impact of LSC gas on measurements of the S-Z towards the central 
Virgo region is appreciable only when observations are made in a 
survey mode - such as planned with PLANCK - with relatively short scan 
time over the region, and consequently insufficient sensitivity to 
remove such an LSC component. Clearly, the additional signal due to LSC 
gas can largely be subtracted out when pointed observations are made 
with a suitably selected beam-through pattern. 

\section{Discussion}

Although there is yet no unequivocal observational evidence for the WHIM, 
theory and hydrodynamic cosmological simulations indicate that it is 
highly prevalent and constitutes a substantial fraction of the baryonic 
mass fraction. Currently available observations of WHIM in several cluster 
regions imply appreciable levels of the Comptonization parameter, 
$\sim 10^{-5}$, comparable to values in most clusters. Even though we 
used a simplified model for the WHIM in the LSC in order to explore some 
of its consequences, we consider the results presented here to be 
qualitatively valid. In fact, our approximation of the WHIM 
filamentary structure by an ellipsoidal configuration (with a large 
volume) quite likely underestimates its S-Z impact, given the basic 
premise that it constitutes an appreciable fraction of the baryon mass 
of the LSC. As indicated by morphological analyses of such filaments 
seen in hydrodynamical simulations, their density grows towards the
central axis, such that a clear preferential direction marked by the
axis can generate even higher S-Z signals along this direction. Moreover, 
the adopted symmetric morphology underestimates the density along the 
main axis in direction to Virgo as there is no cluster in the 
opposite side of the LG-Virgo direction.

Several potential observational consequences of the presence of such 
gaseous component have been discussed in this paper: the angular power 
spectrum of the induced S-Z signal has been calculated and shown to be 
rather low, particularly with regard to what has been speculated in the
literature as a possible explanation for the suppression of primary CMB 
power on the lowest multipoles. While we cannot rule out a suppression 
at the claimed level, we do not find this to be very likely.
The effect of LSC gas on directional S-Z cluster counts is proportional 
to the los crossing the volume of gas in the LSC. On the other hand, a 
definite prediction can be made that the distribution of S-Z cluster 
counts across the sky will be determined to be anisotropic close to the 
level estimated here. The impact is relatively significant owing to the 
highly non-linear form of the mass function; a relatively low S-Z flux 
induced by intervening LSC gas may give rise to a considerable increase 
in the number of potentially detected clusters. This is in particular 
true in light of the decidedly higher abundance of low-flux S-Z clusters 
predicted by the mass function. Needless to say, this effect is most 
noticeable along the main axis towards Virgo. S-Z measurements of the 
Virgo cluster may also be affected by the presence of the WHIM, as well 
as by the dipole distribution of the primary CMB temperature anisotropy. 
This is even more relevant to measurements of the kinematic effect, 
since the induced temperature change does not depend on the temperature 
of the LSC gas, which tends to be lower than typical IC temperatures. 
These effects will have to be taken into account in realistic analyses 
of the S-Z signal from the Virgo cluster. 

Finally, although observational evidence for the WHIM is still scarce,
it is interesting to use the currently available data to assess its
potential of generating a significant S-Z signal towards other SCs. 
Kaastra et al. (2003) list five clusters (A1795, Sersic 159-03, MKW3s, 
A2050, A2199) reported to include a WHIM component, modeled to lie 
within a spherical halo, be isothermal and of homogeneous density. For 
each cluster they list the WHIM electron density, temperature, and size, 
such that the Comptonization parameter along the radius of the spherical 
halo can be easily calculated. They all turn out to lie within the range 
$y\sim (0.4-0.8)\cdot 10^{-5}$, i.e. near the level taken here for the 
LSC. Thus, the S-Z consequences of ISC gas we considered here are likely 
to be relevant in other nearby SCs, but obviously on smaller angular 
scales.

\newpage
\def\ref{\par\noindent\hangindent 20pt}

\end{document}